\documentclass[12pt]{iopart}
\eqnobysec
\def\vep{\varepsilon}

\def\be{\begin{equation}}
\def\ee{\end{equation}}
\def\bea{\begin{eqnarray}}
\def\eea{\end{eqnarray}}

\begin{document}
 \jl{1}
\title{
Squeezing and photon distribution in a vibrating cavity} 
\author{V V Dodonov\footnote{E-mail: vdodonov@power.ufscar.br}\
and M A Andreata}
\address{
Departamento de F\'{\i}sica, Universidade Federal de S\~ao Carlos,\\
Via Washington Luiz km 235, 13565-905  S\~ao Carlos,  SP, Brazil}


\begin{abstract}
We obtain explicit analytical expressions for the quadrature variances
and the photon distribution functions
of the electromagnetic field modes excited
from vacuum or thermal states due to the non-stationary
Casimir effect in an ideal one-dimensional
Fabry--Perot cavity with vibrating walls,
provided the frequency of vibrations is close to a multiple frequency of
the fundamental unperturbed electromagnetic mode.
\end{abstract}

\submitted
\maketitle

\section{Introduction}
During last decade, 
the attention of many authors was attracted to quantum
phenomena in cavities and media with moving boundaries,
known under the names
{\it non-stationary Casimir effect\/} (NSCE) \cite{DKM89},
{\it dynamical Casimir effect\/} \cite{Sch},
or {\it mirror (motion) induced radiation\/} \cite{BE,Lamb}.
For the most recent achievements in this field and references to
other works see, e.g.
[5-23]
(the problems of the {\it classical\/}
electrodynamics with moving boundaries were studied, e.g. in
\cite{Coop,Ditt}).
A complete analytical solution to the problem of
a one-dimensional (Fabry--Perot) ideal cavity with
{\it resonantly vibrating\/} boundaries was found recently in \cite{Djpa}.
This solution holds for any moment of time
(provided the amplitude of the wall vibrations is small enough);
moreover, it enables not only to calculate
the number of photons created from an arbitrary initial state, but also to
account for the effects of detuning from a strict resonance.

In the present article, continuing the line of research of \cite{Djpa},
we calculate the effects of {\it squeezing\/}
and find the {\it photon distribution function\/} of the field in each
mode inside the cavity.
Although a possibility of squeezing the electromagnetic field in a cavity
with moving boundaries due to the NSCE was discussed for the
first time ten years ago \cite{DKM90},
the concrete calculations of the variances of the
field quadrature components were made only in the short-time
\cite{DKM90,CSZ} and long-time \cite{DK92,DKN93} limits under the
condition of the {\it strict\/} resonance.
As to the photon statistics in an oscillating Fabry--Perot cavity,
it was not discussed at all until now
(the photon distribution in a 3D nondegenerate
resonantly driven cavity was found in \cite{DK96}).

\section{Field operator in a 1D cavity with oscillating boundaries}

We start with a brief description of the main results of \cite{Djpa}
which are used in the subsequent sections.
Following the model of a ``scalar electro\-dynamics'' \cite{Moore},
we assume that a 1D cavity is formed by two infinite ideal plates
whose positions are given by
$x_{left}\equiv 0$ and
\be
x_{right}\equiv L(t)=L_0\left(1+\varepsilon \sin\left[p\omega_1
(1+\delta)t\right]\right), \quad |\varepsilon|\ll 1,
\quad |\delta|\ll 1
\label{L(t)}
\ee
where $\omega_1=\pi c/L_0$ and $ p=1,2,\ldots$.
As was shown in \cite{Djpa}, the results can
be easily generalized to a generic case when both the mirrors oscillate
(see also \cite{Ji98,Dal99}).
We are looking for the only component of the operator vector potential
of the electromagnetic field $\hat {A}(x,t)$
{\em in the Heisenberg representation\/} in the form
(hereafter $c=\hbar=1$)
\be
 \hat {A}(x,t)=\sum_{n=1}^{\infty}\frac 2{\sqrt {n}}\left[
\hat b_n\psi^{(n)}(x,t)\,+\,\mbox{h.c.}\,\right],
\quad \left[\hat b_n\,,\,\hat b_k^{\dagger}\right]=\delta_{nk}\, .
\label{vecpot}
\ee
This operator must satisfy the wave equation $\hat{A}_{tt}-\hat{A}_{xx}=0$
and the boundary conditions \cite{Moore}
$\hat{A}(0,t)=\hat{A}(L(t),t)=0$.
For $t<0$, when the wall is assumed to be at rest, function
$\psi^{(n)}(x,t)$ has a simple factorized form
$
\psi^{(n)}_0(x,t)=
e^{-i\omega_nt}
\sin\left(\pi nx/L_0\right)$, $\omega_n =n\omega_1$.
The normalization factors $2/\sqrt{n}$ in (\ref{vecpot}) are
chosen in such a way that the energy of the field in the stationary case
can be represented as a sum of energies of independent mode oscillators
(see the next section).

For $t\ge 0$, following \cite{Law} we expand each function $\psi^{(n)}(x,t)$
in a series with respect to the {\em instantaneous basis\/}
\be
\psi^{(n)}(x,t)=\sqrt {\frac{L_0}{L(t)}}\sum_{k=1}^{\infty}
Q_k^{(n)}(t)
\sin\left[\frac{\pi kx}{L(t)}\right].
\label{decom}
\ee
This way we automatically satisfy the boundary conditions. Then the wave
equation is replaced by the infinite set of coupled
differential equations
\begin{equation}
\ddot {Q}_k^{(n)}+\omega_k^2(t)Q_k^{(n)}
=2\sum_{j=1}^{\infty}  g_{kj}(t)\dot {Q}_j^{(n)}+
\sum_{j=1}^{\infty}  \dot{g}_{kj}(t) Q_j^{(n)}
+{\cal O}\left(g_{kj}^2\right),
\label{Qeq}
\end{equation}
where
\[
\omega_k(t)= {k\pi}/{L(t)}, \quad
g_{kj}=-g_{jk}=(-1)^{k-j}\frac {2kj
\dot {L}(t)}{\left(j^2-k^2\right)L(t)}.
\]
As was shown in \cite{Djpa}, the set of equations (\ref{Qeq}) can be
significantly simplified in the case of the (quasi)resonance oscillations
of the wall given by the dependence (\ref{L(t)}), if one writes
\be
Q_k^{(n)}(t)=
\rho_k^{(n)}e^{-i\omega_k(1+\delta)t}
-\rho_{-k}^{(n)}e^{i\omega_k(1+\delta)t}
\label{decomp}
\ee
assuming the coefficients $\rho_k^{(n)}$
($k=\pm1,\pm2,\ldots$; $n=1,2,\ldots$) to be slowly varying functions of
time, whose derivatives are proportional to the small parameter $\vep$.
Putting (\ref{decomp}) into equation  (\ref{Qeq}) and neglecting
the second order terms like
$\ddot{\rho}_k^{(n)}\sim \varepsilon^2$,
one obtains, after
averaging over fast oscillations with the multiple frequencies of
$\omega_1$, the equations \cite{Djpa}
\begin{equation}
\frac {\mbox{d}}{\mbox{d}\tau}\rho_k^{(n)}=
\sigma\left[(k+p)\rho_{k+p}^{(n)}- (k-p)\rho_{k-p}^{(n)}\right]
+2i\gamma k \rho_{k}^{(n)}
\label{prhok}
\end{equation}
where $\gamma \equiv\delta/\varepsilon$, $\sigma\equiv (-1)^p$, and
$\tau =\frac 12\varepsilon\omega_1t$ is the `slow time'.
The exact solutions to equation (\ref{Qeq}) differ from the approximate
form (\ref{decomp}) by the terms proportional to the higher harmonics
$\exp\left[\pm ir\omega_k t\right]$, $r=2,3,\ldots$. However, the
magnitudes of the corrections are of the order of $\vep$ (or less), so
they can be neglected, at least under the condition
$t\ll t_2\sim \left(\omega_1\varepsilon^2\right)^{-1}$. For the values
$\omega_1\sim 10^{10}$ s$^{-1}$ and $\varepsilon\sim 10^{-8}$
corresponding to the possible experimental realisations \cite{DK96}
the limiting time $t_2$ is of the order of weeks.

Due to the initial conditions $\rho_k^{(n)}(0)=\delta_{kn}$
the solutions to (\ref{prhok}) satisfy the relation
$\rho_{j+mp}^{(k+np)}\equiv 0$  if $j\neq k$.
The non-zero coefficients $\rho_m^{(n)}$ read \cite{Djpa}
\begin{eqnarray}
\rho_{j+mp}^{(j+np)}(\tau)&=&
\frac{\Gamma\left(1+n+j/p\right)
(\sigma\kappa)^{n-m}\lambda^{m+n+2j/p}}
{\Gamma\left(1+m+j/p\right)\Gamma\left(1+n-m\right) }
\nonumber\\
&&\times F\left(n+j/p\,,-m -j/p\,; 1+n-m\,; \kappa^2\right)
\label{rho2}
\end{eqnarray}
where $F(a,b;c;z)$ is the Gauss hypergeometric function,
\be
\kappa = \frac{\sinh(ap\tau)}{\sqrt{a^2+\sinh^2(ap\tau)}}, \quad a =\sqrt{%
1-\gamma^2}, \quad \lambda = \sqrt{1-\gamma^2\kappa^2}+i\gamma\kappa.
\label{defpar}
\ee
The functions (\ref{rho2}) are {\em exact\/} solutions to the set of
equations (\ref{prhok}) relating the coefficients with different lower
indices. Besides, these functions satisfy
another set of equations, which can be treated as
recurrence relations with respect to the {\it upper\/} indices \cite{Djpa}
\begin{equation}
\frac{d }{d\tau}\rho_m^{(n)} = n\left\{\sigma\left[
\rho_m^{(n-p)} -\rho_m^{(n+p)}\right] +2i\gamma \rho_m^{(n)}\right\},
\quad n\ge p, \quad \rho_m^{(0)}\equiv 0
\label{recrho}
\end{equation}
\begin{equation}
\frac{d }{d\tau}\rho_m^{(n)} = n\left\{\sigma\left[
\rho_{-m}^{(p-n)*} -\rho_m^{(p+n)}\right] +2i\gamma \rho_m^{(n)}\right\},
\quad n=1,2,\ldots,p-1
\label{recrho1}
\end{equation}
The consequences of equations (\ref{prhok}), (\ref{recrho}) and
(\ref{recrho1}) are the identities
\begin{eqnarray}
&& \sum_{m=-\infty}^{\infty}
m\rho_{m}^{(n)*}\rho_{m}^{(k)}
=n\delta_{nk}\,, \quad n,k=1,2,\ldots
\label{rhocond1} \\[2mm]
&&
 \sum_{n=1}^{\infty}\frac{m}{n}
\left[\rho_{m}^{(n)*}\rho_{j}^{(n)} -
\rho_{-m}^{(n)*}\rho_{-j}^{(n)} \right] =\delta_{mj}\,,
\quad m,j=1,2,\ldots
\label{rhocond2}\\[2mm]
&&
 \sum_{n=1}^{\infty}\frac{1}{n}
\left[\rho_{m}^{(n)*}\rho_{-j}^{(n)} -
\rho_{j}^{(n)*}\rho_{-m}^{(n)} \right] =0\,,
\quad m,j=1,2,\ldots
\label{rhocond3}
\end{eqnarray}

The formulas given above hold for any value of the detuning parameter
$\gamma$. For $\gamma>1$ one should replace the functions
$\sinh(ax)/a$ and $\cosh(ax)$ by their trigonometrical counterparts
$\sin(\tilde{a}x)/\tilde{a}$ and $\cos(\tilde{a}x)$, where
$\tilde{a}= \sqrt{\gamma^2-1}$.

\section{Squeezing the initial vacuum}

We suppose that after some interval of time $T$
the wall comes back to its initial position $L_0$.
For $t\ge T$, the field operator assumes the form
\be
 \hat {A}(x,t)=\sum_{n=1}^{\infty}\frac 2{\sqrt {n}}
\sin\left(\pi nx/L_0\right)\left[
\hat a_n
e^{-i\omega_n(t+\delta T)}
\,+\,\mbox{h.c.}\,\right]
\label{vecpotfin}
\ee
where operators $\hat {a}_m$ are related to
the initial operators $\hat {b}_n$ and $\hat {b}_n^{\dagger}$
by means of the Bogoliubov transformation
($\tau_T\equiv \frac12 \varepsilon \omega_1 T$)
\begin{equation}
\hat {a}_m=\sum_{n=1}^{\infty}\sqrt {\frac mn} \left[\hat b_n
\rho_m^{(n)}\left(\tau_T\right)
-\hat b_n^{\dag} \rho_{-m}^{(n)*}\left(\tau_T\right)\right],
\quad m=1,2,\ldots .
\label{Bogol}
\end{equation}
The commutation relations
$ \left[\hat a_n\,,\,\hat a_k^{\dagger}\right]=\delta_{nk}$ hold due to
the identities (\ref{rhocond1})-(\ref{rhocond3}) which
are nothing but the {\it unitarity conditions\/} of the transformation
(\ref{Bogol}).
These commutation relations together with the expression for
the energy of the field
\begin{equation}
\hat {H}\equiv\frac 1{8\pi}\int_0^{L_0}\mbox{d}x\,\left
[\left(_{}\frac {\partial \hat{A}}{\partial t}\right)^2+\left(_{}\frac {
\partial \hat{A}}{\partial x}\right)^2\right]
=\sum_{n=1}^{\infty}\omega_n\left(\hat
a^{\dag}_n\hat a_n+\frac 12\right)
\label{Ham}
\end{equation}
convince us that namely $\hat{a}_n$ and $\hat{a}_n^{\dag}$
are true photon annihilation and creation operators at $t\ge T$
(like the operators $\hat{b}_n$ and $\hat{b}_n^{\dag}$ were `physical'
ones at $t<0$).

Now we introduce the Hermitian quadrature component operators
\[
\hat{q}_m=\left(\hat {a}_m +\hat {a}_m^{\dagger}\right)/\sqrt2, \quad
\hat{p}_m=\left(\hat {a}_m -\hat {a}_m^{\dagger}\right)/(i\sqrt2)
\]
and calculate their variances
$
U_m=\langle \hat{q}_m^2\rangle -\langle \hat{q}_m\rangle^2 $,
$V_m=\langle \hat{p}_m^2\rangle -\langle \hat{p}_m\rangle^2$
and the covariance
$Y_m=\frac12\langle \hat{p}_m\hat{q}_m +\hat{q}_m\hat{p}_m\rangle -
\langle \hat{p}_m\rangle\langle \hat{q}_m\rangle$
in the vacuum state {\it with respect to the initial operators\/}
$\hat{b}_n$: $\hat{b}_n |0\rangle=0$ (remember that we use the Heisenberg
picture). Using (\ref{Bogol}) we obtain
(assuming hereafter $\omega_1=1$)
\begin{equation}
U_m= \frac{m}{2}\sum_{n=1}^{\infty} \frac1n
\left|\rho_m^{(n)} -\rho_{-m}^{(n)}\right|^2, \quad
V_m= \frac{m}{2}\sum_{n=1}^{\infty} \frac1n
\left|\rho_m^{(n)} +\rho_{-m}^{(n)}\right|^2
\label{var}
\end{equation}
\begin{equation}
Y_m = \sum_{n=1}^{\infty} \frac{m}{n}
\mathrm{Im} \left[\rho_m^{(n)*}
\rho_{-m}^{(n)}\right]
\label{covar}
\end{equation}
where the coefficients $\rho_{\pm m}^{(n)}$ should be taken at the moment
$T$, thus their argument is $\tau_T$.
Strictly speaking, the expressions (\ref{var})-(\ref{covar})
have physical meanings at those
moments of time $T$ when the wall returns to its initial position,
i.e. for $T=N\pi/[p(1+\delta)]$ with an integer $N$. Consequently,
the argument $\tau_T$ of the coefficients $\rho_{\pm m}^{(n)}$
in (\ref{var})-(\ref{covar}) assumes discrete values
$\tau^{(N)}=N\varepsilon\pi/[2p(1+\delta)]$. One should remember,
however, that something interesting in our problem happens for the values
$\tau\sim 1$ (or larger). Then $N\sim\varepsilon^{-1}\gg 1$, and the
minimal increment $\Delta\tau\sim\varepsilon$ is so small that
$\tau_T$ can be considered as a continuous variable (under the realistic
conditions, $\varepsilon \le 10^{-8}$ \cite{DK96}). For this reason,
we omit hereafter the subscript $T$, writing simply $\tau$ instead of
$\tau_T$ or $\tau^{(N)}$.

Differentiating the right-hand sides of equations (\ref{var}) and
(\ref{covar}) with respect to the `slow time' $\tau$,
one can remove the fraction $1/n$ with the aid of the
recurrence relations (\ref{recrho}) and (\ref{recrho1}).
After that, changing if necessary the summation index $n$ to $n \pm p$,
one can verify that almost all terms in the right-hand sides are cancelled,
and the infinite series are reduced to the finite sums:
\[
\begin{array}{l}
\mbox{d}U_m/\mbox{d}\tau\\
\mbox{d}V_m/\mbox{d}\tau
\end{array}\Bigg\}=
\sigma m\sum_{n=1}^{p-1}{\rm Re}\left(
\left[\rho_{m}^{(p-n)} \mp \rho_{-m}^{(p-n)}\right]
\left[\rho_{-m}^{(n)} \mp \rho_{m}^{(n)}\right]\right)
\]
\[
\mbox{d}Y_m/\mbox{d}\tau=
\sigma m\sum_{n=1}^{p-1} {\rm Im}
\left(\rho_{m}^{(n)*}\rho_{m}^{(p-n)*} +
\rho_{-m}^{(n)}\rho_{-m}^{(p-n)}\right)
\]
Now one should take into account the structure
of the coefficients $\rho_m^{(n)}$  (\ref{rho2}): they are different from
zero provided the difference between the upper index $n$ and the lower one
$m$ is some multiple of the number $p$. If $m=j+pk$
with $j=1,\ldots, p-1$ and $k=0,1,2,\ldots$, then only the terms with
$n=j$ or $n=p-j$ survive in the sums above.
Depending on whether $j=p/2$ or $j \neq p/2$,
we obtain two different sets of explicit expressions for the derivatives
of the (co)variances.

1) If $m=j+pk$
but $j\neq p/2$ (in particular, for all {\it odd\/} values of $p$), then
\begin{equation}
\frac{\mbox{d}U_m}{\mbox{d}\tau}=
\frac{\mbox{d}V_m}{\mbox{d}\tau}=
2\sigma m{\rm Re}\left(\rho_{m}^{(j)}\rho_{-m}^{(p-j)}\right), \quad
\frac{\mbox{d}Y_m}{\mbox{d}\tau}= 0.
\label{pneqj}
\end{equation}
In this case $Y_m\equiv 0$ and $U_m=V_m={\cal N}_m +1/2$,
where ${\cal N}_m$ is the mean number of photons created from vacuum in
the $m$th mode calculated in \cite{Djpa}, so there is no squeezing.
For $\gamma\le 1$ the quadrature variances monotonously
increase in time, with an asymptotical linear dependence.
If $\gamma>1$, the variances oscillate in time
with amplitudes inversly proportional to $\gamma^2 -1$, being always
not less than $1/2$.

2) Some squeezing can be achieved only in the
`principal' modes with the numbers $\mu=p(k+1/2)$, $k=0,1,2,\ldots$:
\begin{equation}
\frac{\mbox{d}U_{\mu}}{\mbox{d}\tau}=
- {\mu}{\rm Re}\left(\left[\rho_{\mu}^{(p/2)}-\rho_{-\mu}^{(p/2)}
\right]^2\right), \quad
\frac{\mbox{d}V_{\mu}}{\mbox{d}\tau}=
 {\mu}{\rm Re}\left(\left[\rho_{\mu}^{(p/2)}+\rho_{-\mu}^{(p/2)}
\right]^2\right)
\label{p2j}
\end{equation}
\begin{equation}
\mbox{d}Y_{\mu}/\mbox{d}\tau = {\mu}\mathrm{Im}\left(
\left[\rho_{\mu}^{(p/2)*}\right]^2
+ \left[\rho_{-\mu}^{(p/2)}\right]^2 \right)
\label{derY}
\end{equation}
(in particular, it is necessary that $p$ be an {\it even\/} number).
In the strict resonance case ($\gamma=0$)
all the coefficients $\rho_{\mu}^{(p/2)}$ are real, so $Y_{\mu}=0$
and $dU_{\mu}/d\tau\le 0$ in the whole interval $0\le \tau<\infty$,
resulting in the inequality $U_{\mu}(\tau)<1/2$.

Note that the coefficients $\rho_{pm+p/2}^{(pn+p/2)}$ depend on the
parameter $p$ only through the dependence of the
the variable $\kappa$ on the product $p\tau$: see
equation (\ref{defpar}) and the explicit form of these coefficients
in  \ref{ap-1}. Thus it is sufficient to consider
the most important special case
of the parametric resonance at the {\it double\/} fundamental frequency
$2\omega_1$ (i.e. $p=2$), since the formulae for $p>2$ can be
obtained by a simple rescaling of the `slow time'
(for the `principal' modes).
In this case, only the odd modes can be excited from
the vacuum, and they do exhibit some squeezing.

Using equations (\ref{p2j}) and (\ref{derY}) one can immediately find
the Taylor expansions of
the (co)variances at $\tau\to 0$
(assuming $(-1)!!\equiv 1$):
\be
\begin{array}{l}
U_{2m+1}\\
V_{2m+1}
\end{array}\Bigg\}
= \frac12 \mp\tau^{2m+1} \left[\frac{(2m-1)!!}{m!}\right]^2
\left[1 \mp\frac{2m+1}{(m+1)^2}\tau +{\cal O}(\tau^2)\right]
\label{UVsmall}
\ee
\be
Y_{2m+1}= - 2\gamma(2m+1)\tau^{2(m+1)} \left[\frac{(2m-1)!!}{m!}\right]^2
+\cdots
\label{Y0}
\ee
We see that the $U$-variances are always less than $1/2$ at the
initial stage, but the degree of their squeezing rapidly decreases
with increase of the number $m$.
Note that the dependence on the detuning parameter $\gamma$ in the
short-time limit appears only in terms of the order of $\tau^{2m+3}$ 
(and higher).

In the opposite limit $\tau\to \infty$ (or $\kappa\to 1$),
using equations (\ref{p2j}), (\ref{derY}) and the asymptotics of the
Bogoliubov coefficients (\ref{F1}) we obtain
{\it constant\/} time derivatives
\bea
\left.\mbox{d}U_{2m+1}/\mbox{d}\tau\right|_{\tau\to\infty}&=&
\frac{16a}{\pi^2(2m+1)}\sin^2\left[\left(m+\frac12\right)\phi\right]
\label{Ubig}\\
\left.\mbox{d}V_{2m+1}/\mbox{d}\tau\right|_{\tau\to\infty}&=&
\frac{16a}{\pi^2(2m+1)}\cos^2\left[\left(m+\frac12\right)\phi\right]
\label{Vbig}\\
\left.\mbox{d}Y_{2m+1}/\mbox{d}\tau\right|_{\tau\to\infty}&=&
-\,\frac{8a}{\pi^2(2m+1)}\sin\left[\left(2m+1\right)\phi\right]
\label{Ybig}
\eea
where $\phi \equiv\arcsin\gamma$. Consequently, all the (co)variances
increase with time linearly, giving the constant photon generation rate
in the `principal' (odd) modes
\be
\left.\mbox{d}{\cal N}_{2m+1}/\mbox{d}\tau
\right|_{\tau\to\infty}
= \frac{8a}{\pi^2(2m+1)}
\label{Ntot}
\ee
 in agreement with \cite{Djpa}.
Equation (\ref{Ntot}) results in a simple estimation of the mean photon
number in the $\mu$th mode at $\tau>1$:
${\cal N}_{\mu}(\tau)\approx a\tau/\mu$.

Since the covariance $Y_{\mu}$ is different from zero if $\gamma\neq 0$,
the initial vacuum state of the field is transformed to the
{\em correlated\/} quantum state \cite{Kur,205}.
One should remember, however, that
the values of $U_{\mu}$, $V_{\mu}$ and $Y_{\mu}$ yield the
(co)variances of the field quadratures only at the moment $t=T$ (when the
wall stopped to oscillate). At the
subsequent moments of time the quadrature variances exhibit fast
oscillations with twice the frequency of the mode. For example (omitting
the mode index),
\[
\sigma_q(t')=U\cos^2(\omega t') +V\sin^2(\omega t')
+Y\sin(2\omega t'), \quad t'=t-T.
\]
Therefore the physical meanings have not the values $U_{\mu}$, $V_{\mu}$
and $Y_{\mu}$ themselves, but rather the {\em minimal\/}
$\sigma_{min}\equiv u_{\mu}$ and
{\em maximal\/} $\sigma_{max}\equiv v_{\mu}$
values of the quadrature variances during the period of fast
oscillations \cite{Pol}
\be
\left.
\begin{array}{l}
u_{\mu}\\
v_{\mu}
\end{array}
\right\}=
 \frac12\left(U_{\mu} +V_{\mu} \mp
\sqrt{\left(U_{\mu} -V_{\mu}\right)^2 +4Y_{\mu}^2}\right).
\label{def-uv}
\ee
Only in the special case of the strict resonane ($\gamma=0$) we have
$u_{\mu}=U_{\mu}$ and $v_{\mu}=V_{\mu}$.

The equations (\ref{p2j}) and (\ref{derY}) can be integrated
for any values of $\mu$ and $\gamma$
in terms of the complete elliptic integrals ${\bf K}(\kappa )$ and
${\bf E}(\kappa )$: see \ref{ap-int} for the technical details.
In particular, for the fundamental mode we find
\[
U_1= \frac 2{\pi^2\kappa}
\left[ \tilde{\kappa}^2(\beta-\kappa){\bf K}^2(\kappa )
-2(\beta-\kappa){\bf K}(\kappa ) {\bf E}(\kappa) +
\beta {\bf E}^2(\kappa)\right],
\]
\[
V_1= \frac 2{\pi^2\kappa}
\left[ 2(\beta+\kappa){\bf K}(\kappa ) {\bf E}(\kappa)
-\tilde{\kappa}^2(\beta+\kappa){\bf K}^2(\kappa ) -
\beta {\bf E}^2(\kappa)\right],
\]
\[
Y_1= \frac {2\gamma}{\pi^2}
\left[
\tilde{\kappa}^2{\bf K}^2(\kappa ) -
2{\bf K}(\kappa ) {\bf E}(\kappa)
+ {\bf E}^2(\kappa)\right]
\]
where $\beta =\mbox{Re}\lambda =\sqrt{1-\gamma^2\kappa^2}$ and
$\tilde{\kappa}\equiv\sqrt{1-\kappa^2}$.
Then equation (\ref{def-uv}) yields
the minimal and maximal invariant variances
\be
u_1= \frac 2{\pi^2\kappa}
\left[ \tilde{\kappa}^2(1-\kappa){\bf K}^2(\kappa )
-2(1-\kappa){\bf K}(\kappa ) {\bf E}(\kappa) +{\bf E}^2(\kappa)\right]
\label{u1min}
\ee
\be
v_1= \frac 2{\pi^2\kappa}
\left[ 2(1+\kappa){\bf K}(\kappa ) {\bf E}(\kappa)
-\tilde{\kappa}^2(1+\kappa){\bf K}^2(\kappa ) - {\bf E}^2(\kappa)\right]
\label{v1min}
\ee
which depend on the detuning
parameter $\gamma$ only implicitly, through the dependence on $\gamma$
of the function $\kappa(\tau)$ (\ref{defpar}).
In the short time limit $\tau\ll 1$ (then $\kappa\approx 2\tau$) we obtain,
using the Taylor expansions of the complete elliptic integrals,
$u_1=\frac12 -\tau +\tau^2 +\cdots$ and
$v_1=\frac12 +\tau +\tau^2 +\cdots$
in accordance with \cite{DKM90}.
More precisely,
\[
\begin{array}{l}
u_1\\
v_1
\end{array}\Bigg\}
=\frac12\left( 1 \mp \kappa +\frac12\kappa^2 \mp \frac14\kappa^3
+\frac{7}{32}\kappa^4 +\cdots\right)
\]
The minimal variance $u_1$ monotonously decreases from the value $1/2$
at $t=0$ to the constant asymptotical value $2/\pi^2$ at $\tau\gg 1$,
confirming qualitatively the evaluations of \cite{DK92,DKN93}
and giving almost 50\% squeezing in the initial vacuum state.
The variance of the conjugate quadrature monotonously increases,
and for $\tau\gg 1$ it becomes practically linear function of time:
$v_1(\tau\gg 1)\approx 16\tau/\pi^2$.  The asymptotical minimal value
$u_1(\tau=\infty)$ does not depend on $\gamma$ provided $\gamma\le 1$
(only the rate of reaching this asymptotical value decreases with $\gamma$
as $\sqrt{1-\gamma^2}$). In the strongly detuned case, $\gamma >1$,
the minimal variance oscillates as a function of $\tau$
(being always greater than $2/\pi^2$),
since in this case the function $\kappa(\tau)$ oscillates between
$-\gamma^{-1}$ and $\gamma^{-1}$.

The minimal variance does not go to zero when $\tau\to\infty$ due to the
{\em strong intermode interaction\/}, which results in a high degree
of {\em quantum mixing\/} for each mode.
Since the state originating from the initial
vacuum state belongs to the class of {\em Gaussian\/} states
(see section \ref{PDF}), the quantum `purity'
$\chi_m\equiv\mbox{Tr}\hat\rho_m^2$
of the $m$th field mode (described by means of the density matrix
$\hat\rho_m$) can be expressed in
terms of the (co)variances as \cite{167}
$\chi_m=\left[4\left(U_m V_m -Y_m^2\right)\right]^{-1/2}$.
Using equations (\ref{Ubig})-(\ref{Ybig}) one can check that
 $\chi \sim \tau^{-1/2} \to 0$ for $\tau\gg 1$ (see \ref{ap-2}).
For instance, for $m=1$ we have (writing simply ${\bf K}$ and ${\bf E}$
instead of ${\bf K}(\kappa)$ and ${\bf E}(\kappa)$)
\be
\chi_1=\frac{\pi^2}{4}\kappa\left[4{\bf K}{\bf E}^3
+4\tilde{\kappa}^4{\bf K}^3{\bf E}
-6\tilde{\kappa}^2{\bf K}^2{\bf E}^2
-{\bf E}^4
-\tilde{\kappa}^6{\bf K}^4 \right]^{-1/2}
\label{chi}
\ee
The initial dependence on $\kappa$ is rather weak:
$\chi(\kappa\ll 1) = 1-\frac{3}{32}\kappa^4 +\cdots$.
But when $\kappa\to 1$, $\chi$ rapidly goes to zero:
$\chi(\tilde{\kappa}\ll 1)\approx (8/\pi^2)
\left[\ln\left(4/\tilde{\kappa}\right)\right]^{-1/2}$, with
$\mbox{d}\chi/\mbox{d}\kappa \to -\infty$.

It is worth to mention that in the resonance case there is no effective
interaction between different modes of a
{\em nondegenerate\/} (three-dimensional) cavity possessing
non-equidistant eigenmode spectrum. In such a cavity
the minimal variance asymptotically goes to zero for $\tau\to\infty$
\cite{DK96,D98}.

For the sake of completeness we bring the formulae
(found in \cite{Djpa}) for the mean
photon number in the fundamental mode and the total energy of photons
created in all the modes (for the initial vacuum state):
\begin{equation}
{\cal N}_1(\kappa )=
\frac 2{\pi^2}{\bf K}(\kappa )
\left[2{\bf E}(\kappa) -\tilde{\kappa}^2{\bf K}(\kappa )\right]
-\frac 12
\label{num1EK}
\end{equation}
\begin{equation}
{\cal E}_{tot}(\tau)= \frac{p^2-1}{12a^2}\sinh^2(pa\tau)\,.
\label{Etotvac}
\end{equation}
In \cite{Djpa} it was derived also the formula for the second derivative
of the total mean number of photons created from vacuum in all the modes
\be
\frac {\mbox{d}^2{\cal N}_{tot}}{\mbox{d}\tau^2}=
\frac 8{\pi^2\kappa^2}
\left[\tilde{\kappa}^4{\bf K}^2
-2\tilde{\kappa}^2{\bf K}{\bf E}
+\left(1+\kappa^2 -2\gamma^2\kappa^4\right){\bf E}^2\right].
\label{secder}
\ee
Now we are able to integrate this equation. Taking
into account the condition \cite{Djpa}
$\mbox{d}{\cal N}/\mbox{d}\tau=0$ at $\tau=0$
we obtain very simple expression
\be
{\cal N}_{tot}=
\frac 2{\pi^2} {\bf K}(\kappa)
\left[{\bf K}(\kappa) -{\bf E}(\kappa) \right].
\label{intsecder}
\ee

The expressions for the variances in the modes with numbers $\mu\ge 3$
are rather involved;
their general structure is discussed in \ref{ap-int}.
Here we give only one explicit example -- the variance $U_3$ for $\gamma=0$:
\bea
U_3 &=& \frac 2{9\pi^2\kappa^3}
\left[ \tilde{\kappa}^2(1-\kappa)
\left( 4 +10\kappa +9\kappa^2\right){\bf K}^2(\kappa )\right. \nonumber\\
&&\left.+(1-\kappa)\left( 4\kappa^3 -14\kappa^2 -20\kappa -8\right)
{\bf K}(\kappa ) {\bf E}(\kappa) \right. \nonumber\\
&&\left.+\left( 4\kappa^4 +6\kappa^3 -\kappa^2 +6\kappa +4\right)
{\bf E}^2(\kappa)\right]
\label{U3}
\eea
The Taylor expansion of the right-hand side of
(\ref{U3}) coincides with the expansion (\ref{UVsmall}). The asymptotical
value at $\tau\to\infty$ equals
$U_3(\kappa=1) =38/(9\pi^2)\approx 0.43$. We see that the squeezing
rapidly disappears with increase of the mode number $\mu$.

The variance $V_3$ can be obtained from (\ref{U3}) by means of a simple
substitution $\kappa\rightarrow -\kappa$. Therefore the mean number of
photons in the third mode is given by
\begin{equation}
{\cal N}_3=
\frac 2{3\pi^2\kappa^2}\left[\left(3\kappa^2 -2\right){\bf K}
\left(2{\bf E} -\tilde{\kappa}^2{\bf K}\right)
+2\left(1+\kappa^2\right){\bf E}^2\right]
-\frac 12\,.
\label{num3}
\end{equation}
It is remarkable that despite all three (co)variances $U_{\mu}$,
$V_{\mu}$ and $Y_{\mu}$ linearly increase with time at $\tau\gg 1$ in
the generic case $\gamma\neq 0$, the {\em minimal variance\/} $u_{\mu}$
tends to a {\em constant\/} value at $\tau\to\infty$: see \ref{ap-2}.

The results of this section confirm completely
the earlier conclusions \cite{DKM90}
concerning the behaviour of the quadrature variances in the short-time
limit for $p=2$, as well as the results of
approximate numerical calculations performed in \cite{CSZ} (in the same
limit) for $p=2$ and $p=3$.
In the long-time limit we see a {\it qualitative\/} agreement with earlier
approximate asymptotical formulae of \cite{DK92,DKN93} related to the
behaviour of the squeezed quadrature component variance (for $\gamma=0$),
namely, that the squeezing effect is strongest for the lowest mode and that
it decreases with increase of the mode number. However, there are
differences in the numerical values of the squeezed variances. This can be
explained as follows. In \cite{DK92,DKN93} only the
{\it leading terms\/} of the
Bogoliubov coefficients (analogs of the coefficients $\rho_m^{(n)}$) were
found. But the coefficients
$\rho_m^{(n)}(\kappa)$ and $\rho_{-m}^{(n)}(\kappa)$ become the same in
the limit $\kappa=1$ (if $\lambda=1$)
and they do not depend on the upper index $n$: see \ref{ap-1}.
For this reason, although the leading terms of the asymptotic
expansions enable to
calculate correctly the number of photons and the {\it unsqueezed\/}
variances like $V_{2m+1}$, these terms are
{\it canceled\/} in the expressions for the {\it squeezed\/} variances
like $U_{2m+1}$ (if $\gamma=0$). It was found in
\cite{DK92,DKN93} that the difference
$\Delta_m\equiv\frac12 - U_{2m+1}(\tau\to\infty)$ decreases
as $1/(2m+1)$ for $m\gg 1$.
The numerical integration of equation (\ref{p2j}) yields for the
product $(2m+1)\Delta_m$ the values close to $0.2$
for $1\le m\le 5$
(according to (\ref{U3}), $3\Delta_1=0.2166\ldots$).
It seems probable that the limit value of this
product at $m\to\infty$ equals $2/\pi^2=0.2026\ldots$,
but we did not succeed to prove this conjecture analytically.

\section{Influence of initial conditions}\label{ap-sqab}

For an arbitrary initial state of the field one can write
 $U_m=U_m^{(vac)} +\Delta U_m$, where $U_m^{(vac)}$ is given by
equation (\ref{var}); similar expressions can be written for $V_m$ and $Y_m$.
The corrections due to the nonvacuum initial states are given by
\bea
\begin{array}{l}
\Delta U_m \\
\Delta V_m
\end{array}
\Bigg\} &=&
\mathrm{Re} \sum_{n,j} \frac{m}{\sqrt{nj}}\Bigg(
\left[\rho_m^{(n)} \mp \rho_{-m}^{(n)}\right]^*
\left[\rho_m^{(j)} \mp \rho_{-m}^{(j)}\right]
\left[\langle \hat{b}_n^{\dag} \hat{b}_j \rangle
-\langle \hat{b}_n^{\dag} \rangle\langle\hat{b}_j \rangle \right]
 \nonumber\\
&& \pm
\left[\rho_m^{(n)} \mp \rho_{-m}^{(n)}\right]
\left[\rho_m^{(j)} \mp \rho_{-m}^{(j)}\right]
\left[\langle \hat{b}_n \hat{b}_j \rangle
-\langle \hat{b}_n \rangle\langle\hat{b}_j \rangle \right]\Bigg)
\label{deltavar}
\eea
\bea
\Delta Y_m  &=&
\mathrm{Im} \sum_{n,j} \frac{m}{\sqrt{nj}}\Bigg(
\left[\rho_m^{(n)*} \rho_{-m}^{(j)} - \rho_m^{(j)}\rho_{-m}^{(n)*}\right]
\left[\langle \hat{b}_n^{\dag} \hat{b}_j \rangle
-\langle \hat{b}_n^{\dag} \rangle\langle\hat{b}_j \rangle \right]
 \nonumber\\
&& +
\left[\rho_m^{(n)} \rho_m^{(j)} - \rho_{-m}^{(n)} \rho_{-m}^{(j)}\right]
\left[\langle \hat{b}_n \hat{b}_j \rangle
-\langle \hat{b}_n \rangle\langle\hat{b}_j \rangle \right]\Bigg)
\label{deltacovar}
\eea
where the average values like $\langle\hat{b}_n^{\dag} \hat{b}_j \rangle$
are calculated in the initial state.
All the corrections disappear in the case of the initial
coherent state,
$\hat{b}_n \vert \alpha\rangle =\alpha_n \vert \alpha\rangle$.
If the initial density matrix is {\em diagonal\/} in the Fock basis
(as happens e.g. for the Fock or thermal states)
then $\langle \hat{b}_n^{\dag} \hat{b}_j \rangle =\nu_n \delta_{nj}$
($\nu_n \ge 0$),
all other average values in (\ref{deltavar}) and (\ref{deltacovar})
being equal to zero. In this case the double sums are reduced to the
single ones:
\be
\Delta U_m =
m \sum_{n} \frac{\nu_n}{n}
\left|\rho_m^{(n)} - \rho_{-m}^{(n)}\right|^2 , \quad
\Delta V_m =
m \sum_{n} \frac{\nu_n}{n}
\left|\rho_m^{(n)} + \rho_{-m}^{(n)}\right|^2
\label{vartherm}
\ee
\be
\Delta Y_m =
2m \sum_{n} \frac{\nu_n}{n}
\mathrm{Im}\left[\rho_m^{(n)*} \rho_{-m}^{(n)}\right].
\label{covartherm}
\ee
We see that the initial fluctuations always increase both the
variances $U_m$ and $V_m$ (for the diagonal density matrix).
However, asymptotically at $\tau\to\infty$
the corrections are bounded for the {\em physical\/} initial
states having finite total numbers of photons,
because the coefficients
$\left|\rho_m^{(n)} \pm \rho_{-m}^{(n)}\right|^2$ and
$\mathrm{Im}\left[\rho_m^{(n)*} \rho_{-m}^{(n)}\right]$ do not depend on the
summation index $n$ in this limit:
see equations (\ref{asrho+}) and (\ref{asrho-}). Thus we have
 in the `principal' $\mu$-modes (for $p=2$; $\phi \equiv\arcsin\gamma$)
\be
\left.
\begin{array}{l}
\Delta U_{\mu}^{(\infty)}\\
\Delta V_{\mu}^{(\infty)}\\
\Delta Y_{\mu}^{(\infty)}
\end{array}\right\}=
\frac{8{\cal Z}}{\pi^2\mu}\times \left\{
\begin{array}{l}
2\sin^2\left(\mu\phi/2\right)\\
2\cos^2\left(\mu\phi/2\right)\\
-\sin\left(\mu\phi\right)
\end{array},\right.
 \quad
{\cal Z}=\sum_{k=0}^{\infty} \frac{\nu_{2k+1}}{2k+1} .
\label{delUinf}
\ee
These expressions are very similar to (\ref{Ubig})-(\ref{Ybig}).
Their consequence is the important result that in the limit $\tau\to\infty$
the minimal variance $u_{\mu}$ {\em does not depend\/} on the initial state
of the field inside the cavity,
provided the initial density matrix was diagonal in the Fock basis. The proof
is given in \ref{ap-2}. The correction to the mean number of photons
tends to the limit
$\Delta {\cal N}_{\mu}^{(\infty)}= 8{\cal Z}/\left(\pi^2\mu\right)$.

\section{Photon distribution}\label{PDF}

Now let us turn to the {\it photon distribution function\/} (PDF)
$
f({\bf n})\equiv \langle {\bf n}|\hat\rho_m(t)|{\bf n}\rangle
$,
where $|{\bf n}\rangle$ is the multimode Fock state,  
${\bf n}\equiv\left(n_1,n_2,\ldots\right)$, and $\hat\rho_m(t)$ is
the time-dependent density matrix of the $m$th field mode in the
{\it Schr\"odinger\/} picture.
At first glance, there is a problem, since all the calculations in the
preceding sections were performed in the framework of the
{\it Heisenberg\/} picture.
Fortunately, this problem can be easily resolved for a special
class of {\it Gaussian\/} initial states (i.e. the states whose
density matrices, or wave functions, or Wigner functions, are
described by some Gaussian exponentials). This class includes
coherent, squeezed and thermal states; in particular, it includes
the vacuum state which we are interested in here.

The solution is based on two key points. The first one is
the statement \cite{Plun,Law}
that the field evolution in a cavity with moving boundaries
can be described not only in the Heisenberg picture, but, 
equivalently, in the
framework of the Schr\"odinger picture, with a {\it quadratic\/}
multidimensional time-dependent Hamiltonian. 
The second key point is the fact \cite{75,183} 
that the evolution governed by quadratic Hamiltonians
transforms any Gaussian state to another Gaussian state. 

It remains to take into account that the photon distribution function of
any Gaussian state is determined completely by the
average values of quadratures and by their variances
\cite{1mod,DM94},
which obviously do not depend on the quantum mechanical
 representation.
The explicit formulae in the generic case are rather involved, so we give
them in \ref{ap-3}. Here we confine ourselves to the most simple case of
the {\em vacuum\/} initial state, when
all the average values of quadratures are equal to zero.
In this case the generating function (\ref{GF}) is reduced to
$[{\cal G}(z)]^{-1/2}$, i.e. it has the same structure
as the known generating function of the Legendre polynomials $P_n(x)$.
After some algebra we obtain the following expression for the
photon distribution in the $m$th field mode:
\begin{equation}
f_m(n)= \frac{2\left[(2u_m-1)(2v_m-1)\right]^{n/2}}
{\left[(2u_m+1)(2v_m+1)\right]^{(n+1)/2}}
P_n\left(\frac {4u_m v_m- 1}{
\sqrt{\left(4u_m^2-1\right)\left(4v_m^2-1\right)}}\right)
\label{fotdis}
\end{equation}
It depends only on the invariant minimal and maximal variances
$u_m$ and $v_m$.
Note that the argument of the polynomial in (\ref{fotdis}) is always
{\it outside\/} the `traditional' interval $(-1,1)$ 
(in particular, this argument is pure imaginary if $2u_m <1$), being exactly
equal to $1$ for the `non-principal' modes with
$u_m=v_m={\cal N}_m +\frac12$,
when formula (\ref{fotdis}) transforms to the
 time-dependent Planck's distribution
\[
f_m(n;\tau)=\frac{{\cal N}_m^n(\tau)}
{\left[{\cal N}_m(\tau)+1\right]^{n+1}}.
\]
Only for the `principal' $\mu$-modes the spectrum of photons is
different from Planck's one due to the squeezing effect. 
The first and second derivatives of the generating function (\ref{GF})
at $z=1$ yield the first two moments of the photon distribution
(hereafter we suppress subscript $m$)
\be
\bar{n} = \frac12(u+v-1), \quad
\sigma_n\equiv \overline{n^2} -(\bar{n})^2=
\frac14\left(2u^2 +2v^2  -1\right)
\label{var-n}
\ee
which result in the Mandel parameter
\be
Q\equiv \sigma_{n}/\bar{n} -1=
\frac{u^2 +v^2  - u -v + 1/2}{u+v-1}.
\label{Mand}
\ee
This parameter appears positive for all values of $\tau$, so the photon
statistics is super-Poissonian, with strong bunching of photons
(the pair creation of photons in the NSCE was discussed in
\cite{BE,Lamb,Mun,Lamb98}).
In particular,
\[
Q_{2m+1}(\tau\to 0)\approx \left[(m+1)(2m-1)!!/m!\right]^2 \tau^{2m}/(2m+1),
\quad Q_1(0)=1,
\]
whereas $Q_m\approx V_m(\tau)\gg 1$ for $\tau\gg 1$  (if $\gamma\ll 1$).

The analytical expressions obtained in the paper are illustrated in two
figures. In figure~\ref{fig-1} we show
the dependences on the universal variable $\kappa$ (\ref{defpar})
of the minimal and maximal variances
$u_1$ (\ref{u1min}) and $v_1$ (\ref{v1min}) together with the
purity factor $\chi_1$ (\ref{chi}), the mean photon number
${\cal N}_1$ (\ref{num1EK})  and
the $Q$-factor (\ref{Mand}) for the fundamental mode $\mu=1$.
An example of the photon distribution in the principal mode $\mu=1$
for $\gamma=0$ is given in figure~\ref{fig-comp}, where the `cavity'
distribution is compared with Planck's one corresponding to the same
mean photon number. We see no oscillations typical for the pure squeezed
states [40-42]
(excluding a small `splash' at $n=2$),
because the field appears in a {\it mixed\/} quantum state
(the influence of quantum
mixing on the oscillations of the PDF in generic Gaussian states
was studied in \cite{1mod,DM94,D94}).
The asymptotics of the photon distribution function (\ref{fotdis}) in
the long-time limit are given in \ref{ap-3}.

\section{Conclusion}

The main results of the paper are as follows. We studied the behavior of
the electromagnetic field quadrature variances in a one-dimensional
cavity with resonantly oscillating ideal boundaries
in the whole time interval $0\le t<\infty$ for
all field modes and for any (small) value of the detuning parameter.
We have shown that each field mode goes to
a mixed quantum state due to the intermode interaction (caused by
Doppler's effect on the moving mirrors).
We found that squeezing can be observed only in the `principal' modes
with numbers $p(k+1/2)$, where integer $p$ is close to the ratio of the
wall vibration
frequency to the frequency of the fundamental cavity mode.
Analyzing the influence of the
initial nonvacuum state of the field
we discovered that the initial thermal fluctuations do not affect
the minimal value of the quadrature variance (which is less than $1/2$)
in the long-time limit.
This result is important from the practical point of view, since it shows
that certain significant features of the Nonstationary Casimir Effect are
not sensitive to the temperature (see also \cite{Plun-temp} in the case of
the three-dimensional cavity).
We found the photon
distribution functions $f_m(n)$ for all modes.
For the modes which do not exhibit squeezing,
the PDF coincides with a time-dependent Planck's distribution,
while the PDF in the distinct `principal' modes
differs from Planck's one, being more `flat' for $n\gg 1$.

\section*{Acknowledgement}

MAA thanks Brazilian agency CNPq for the support (project 110524/97-7).

\newpage
\appendix

\section{Some properties of the Bogoliubov coefficients $\rho_m^{(n)}$}
\label{ap-1}

The non-zero coefficients $\rho_{\mu}^{(n)}$ for the `principal'
modes read
\begin{eqnarray}
\rho_{pm+p/2}^{(pn+p/2)}&=&
\frac{\Gamma\left(n+3/2\right)
\kappa^{n-m}\lambda^{m+n+1}}
{\Gamma\left(m+3/2\right)\Gamma\left(1+n-m\right) }
\nonumber\\
&\times&
F\left(n+1/2\,,\,-m -1/2\,;\, 1+n-m\,;\, \kappa^2\right),
\quad n\ge m
\label{xinm}
\end{eqnarray}
\begin{eqnarray}
\rho_{pm+p/2}^{(pn+p/2)}&=&
\frac{(-1)^{m-n} \Gamma\left(m+1/2\right)
\kappa^{m-n}\lambda^{m+n+1}}
{\Gamma\left(n+1/2\right) \Gamma\left(1+m-n\right) }
\nonumber\\
&\times&
F\left(m+1/2\,,\,-n -1/2\,;\, 1+m-n\,;\, \kappa^2\right),
\quad m\ge n
\label{ximn}
\end{eqnarray}
\begin{eqnarray}
\rho_{-pm-p/2}^{(pn+p/2)}&=&
\frac{(-1)^{m}\Gamma\left(m+1/2\right)\Gamma\left(n+3/2\right)
\kappa^{n+m+1}\lambda^{n-m} }
{\pi\Gamma\left(2+n+m\right) }
\nonumber\\
&\times&
F\left(n+1/2\,,\,m +1/2\,;\, 2+n+m\,;\, \kappa^2\right).
\label{etank}
\end{eqnarray}
Using the formula \cite{Grad}
\be
F(a,b;a+b+1;1)=\frac{\Gamma(a+b+1)}{\Gamma(a+1)\Gamma(b+1)}
\label{F1}
\ee
one can find the asymptotics of the coefficients
$\rho_{m}^{(n)}$ for $\kappa\to 1$ \cite{Djpa}
\be
\rho_{pm+j}^{(pn+j)}(\tau\gg 1) \approx \frac{\sin[\pi(m+j/p)]}{\pi(m+j/p)}
(a+i\gamma)^{m+n+2j/p}\sigma^{n-m}
\label{asy}
\ee
In particular, for $p=2$ and for the odd (`principal') modes
\be
\rho_{2m+1}^{(2n+1)}(\tau\gg 1) \approx \frac{2(-1)^m}{\pi(2m+1)}
(a+i\gamma)^{m+n+1},
\label{asrho+}
\ee
\be
\rho_{-2m-1}^{(2n+1)}(\tau\gg 1) \approx \frac{2(-1)^m}{\pi(2m+1)}
(a+i\gamma)^{n-m}
\label{asrho-}
\ee

It is known \cite{BrMar} that the hypergeometric function $F(a,b;c;z)$
with `half-integral' parameters $a,b$ and an integral parameter $c$ can be
expressed in terms of the complete elliptic integrals
\be
{\bf K}(\kappa )=\int_0^{\pi /2}\frac {\mbox{d}\alpha}{\sqrt {1
-\kappa^2\sin^2\alpha}}
=\frac{\pi}{2} F\left(\frac12\,,\,\frac12\,;\,1\,;\,\kappa^2\right)
\label{defK}
\ee
\be
{\bf E}(\kappa )=\int_0^{\pi /2}\mbox{d}\alpha
\sqrt {1-\kappa^2\sin^2\alpha}
=\frac{\pi}{2} F\left(-\frac12\,,\,\frac12\,;\,1\,;\,\kappa^2\right).
\label{defE}
\ee
In particular,
\begin{equation}
\rho_1^{(1)}=\frac{2\lambda(\kappa)}{\pi}{\bf E}(\kappa), \quad
\rho_{-1}^{(1)}=\frac{2}{\pi\kappa}\left[{\bf E}(\kappa )
-\tilde{\kappa}^2{\bf K}(\kappa ) \right]
\label{xiet1}
\end{equation}
\begin{equation}
\rho_3^{(1)}=
\frac{2\lambda^2(\kappa)}{3\pi\kappa}\left[\left(1-2\kappa^2\right)
{\bf E}(\kappa) -\tilde{\kappa}^2{\bf K}(\kappa ) \right]
\label{xi3}
\ee
\be
\rho_{-3}^{(1)}=
-\,\frac{2}{3\pi\kappa^2\lambda(\kappa)}\left[\left(2-\kappa^2\right)
{\bf E}(\kappa) -2\tilde{\kappa}^2{\bf K}(\kappa ) \right]
\label{et3}
\end{equation}
where $\tilde{\kappa}\equiv\sqrt{1-\kappa^2}$ and
$\lambda(\kappa)=\sqrt{1-\gamma^2\kappa^2} +i\gamma\kappa$.

The general structure of the coefficients $\rho_{\mu}^{(1)}$
(we confine ourselves to the case $p=2$) is as follows
\begin{equation}
\rho_{2m+1}^{(1)}=
\frac{2\lambda^{m+1}(\kappa)}{\pi\kappa^m}\left[
f_m\left(\kappa^2\right) {\bf E}(\kappa)
+\tilde{\kappa}^2 g_m\left(\kappa^2\right) {\bf K}(\kappa ) \right]
\label{xim}
\ee
\be
\rho_{-2m-1}^{(1)}=
\frac{2}{\pi\kappa^{m+1}\lambda^m(\kappa)}\left[
r_m\left(\kappa^2\right) {\bf E}(\kappa)
+\tilde{\kappa}^2 s_m\left(\kappa^2\right){\bf K}(\kappa ) \right]
\label{etm}
\end{equation}
where $f_m(x), g_m(x), r_m(x), s_m(x)$ are the polynomials of the degree $m$
which can be found from the recurrence relations (\ref{prhok}).

The Taylor expansions of the complete elliptic integrals at $\kappa\to 0$
(when $\kappa\approx ap\tau$) read
\[
{\bf E}(\kappa)=\frac{\pi}{2}
\left(1-\frac14\kappa^2 -\frac{3}{64}\kappa^4 +\cdots\right), \quad
{\bf K}(\kappa)=\frac{\pi}{2}
\left(1+\frac14\kappa^2 +\frac{9}{64}\kappa^4 +\cdots\right)
\]
whereas their asymptotic behaviours at $\tilde{\kappa}\to 0$ are given by
the formulae \cite{Grad}
\begin{eqnarray*}
{\bf K}(\kappa )&\approx&\ln\frac 4{\tilde{\kappa}}
+\frac 14\left(\ln\frac 4{\tilde{\kappa}}-1\right)\tilde{\kappa}^
2+\cdots \\
{\bf E}(\kappa )&\approx& 1+\frac 12\left(\ln\frac
4{\tilde{\kappa}}-\frac 12\right)\tilde{\kappa}^2 +\cdots \;.
\end{eqnarray*}
In this case $\tilde{\kappa}\approx a/\sinh(ap\tau)$ and
$\ln( 1/\tilde{\kappa}) \approx ap\tau$.

\section{Calculation of integrals}\label{ap-int}

To calculate, for instance, the variance $U_1$ we use equations
(\ref{p2j}) and (\ref{xiet1}) and replace the derivative over $\tau$
by the derivative with respect to $\kappa$ using the relation (if $p=2$)
$\mbox{d}\kappa=2\beta\tilde{\kappa}^2\mbox{d}\tau$, where
$\beta=\sqrt{1-\gamma^2\kappa^2}$. We arrive at the equation
\bea
\frac{\mbox{d}U_1}{\mbox{d}\kappa}&=& -\,
\frac2{\pi^2 \tilde{\kappa}^2 \kappa^2 \beta} \left\{
\left[\kappa^2\left(1-2\gamma^2\kappa^2\right) +1 -2\beta\kappa\right]
{\bf E}^2(\kappa )
\right.\nonumber\\  && \left.
 - 2\tilde{\kappa}^2(1 -\beta\kappa)
{\bf E}(\kappa ){\bf K}(\kappa ) + \tilde{\kappa}^4 {\bf K}^2(\kappa )
\right\}.
\label{eqU1}
\eea
Let us consider first the case $\gamma=0$, when $\beta=1$.
Taking into account the differentiation rules \cite{Grad}
\begin{equation}
\frac {\mbox{d}{\bf K}(\kappa )}{\mbox{d}\kappa}=\frac {
{\bf E}(\kappa )}{\kappa\tilde{\kappa}^2}-\frac {{\bf K}(\kappa )}{
\kappa},\quad
\frac {\mbox{d}{\bf E}(\kappa )}{\mbox{d}\kappa}=\frac {
{\bf E}(\kappa )-{\bf K}(\kappa )}{\kappa}
\label{difrul}
\end{equation}
we may suppose that the factor $\tilde{\kappa}^2$ in the denominator
of the right-hand side of equation (\ref{eqU1}) comes from the derivative
$\mbox{d}{\bf K}/\mbox{d}\kappa $. Thus it is natural to look for the
solution in the form
\be
U_1= \frac 2{\pi^2\kappa}
\left[ A(\kappa){\bf K}^2(\kappa )
+B(\kappa){\bf K}(\kappa ) {\bf E}(\kappa) +
C(\kappa) {\bf E}^2(\kappa)\right],
\label{try}
\ee
where $A(\kappa)$, $B(\kappa)$ and $C(\kappa)$ are some polynomials of
$\kappa$. Putting the expression (\ref{try}) into equation (\ref{eqU1})
we obtain a set of coupled equations for the unknown functions
$A,B,C$. Writing $A(\kappa)=a_0 +A_1(\kappa)$,
$B(\kappa)=b_0 +B_1(\kappa)$, $C(\kappa)=c_0 +C_1(\kappa)$
we determine the constant coefficients
$a_0$, $b_0$ and $c_0$ by putting $\kappa=0$ in that equations.
Then we obtain new
equations for the functions $A_1(\kappa)$, $B_1(\kappa)$ and $C_1(\kappa)$
and repeat the procedure. After a few steps we arrive at the equations
which have obvious trivial solutions $A_n=B_n=C_n=0$. This confirms our
hypothesis on the polynomial structure of the functions
$A(\kappa)$, $B(\kappa)$ and $C(\kappa)$ and gives the final answer.
The equations for the variances $U_{\mu}$, $V_{\mu}$, etc. with $\mu\ge 3$
can be integrated in the same manner,
the only difference is that one should write $\kappa^{\mu}$ instead of
$\kappa$ in the denominator of the expression like (\ref{try}).
In the generic case $\gamma\neq 0$ we notice that the factor $\beta$
can appear in the denominator of the expression (\ref{eqU1}) as a result
of differentiating the function $\beta(\kappa)$, since
$\mbox{d}\beta/\mbox{d}\kappa= -\gamma^2\kappa/\beta$. Therefore we split
each function, $A,B,C$ in the `$\beta$-even' and `$\beta$-odd' parts like
$A=A_e(\kappa) +\beta(\kappa) A_o(\kappa)$.
The equations for the `even' and `odd'
coefficients turn out independent, and we solve them using the procedure
described above.
The equation (\ref{secder}) was integrated using the same scheme.

\section{Asymptotics of the minimal variance and purity factor}\label{ap-2}

For the initial diagonal density matrix (in the Fock basis),
combining the equations (\ref{Ubig})-(\ref{Ybig}) and
(\ref{delUinf}), we write the variances at $\tau\gg 1$ as
(we omit the subscript $\mu$)
\be
\left(
\begin{array}{l}
U(\tau)\\
V(\tau)\\
Y(\tau)
\end{array}\right)=
\left(
\begin{array}{l}
2F\sin^2(\chi/2) + f\\
2F\cos^2(\chi/2) + g\\
-F\sin\chi+ h
\end{array}\right), \quad
F=\frac{8(a\tau +{\cal Z})}{\pi^2\mu}, \quad \chi=\mu\phi.
\label{vartherminf}
\ee
The functions $f$, $g$ and $h$ are much smaller than $F$.
At $\tau\to\infty$ these functions tend to finite limits
which do not depend on the initial state, since they can be found
by integrating equations (\ref{Ubig})-(\ref{Ybig}).
Thus we have $U +V= 2F +f +g$, whereas
\[
(U -V)^2 +4Y^2=
4F^2 +4F[(g-f)\cos\chi -2h\sin\chi] +(f-g)^2 +4h^2 .
\]
For $F\gg f,g,h$ we have
\[
\sqrt{(U -V)^2 +4Y^2} = 2F + (g-f)\cos\chi -2h\sin\chi +{\cal O}(1/F)
\]
so the minimal variance $u(\tau)$ (\ref{def-uv}) tends to the finite limit
\[
u(\infty)= f\cos^2(\chi/2) + g\sin^2(\chi/2) +h\sin\chi
\]
which does not depend on ${\cal Z}$, i.e. on the initial state.

Analogously, $UV-Y^2 =2Fu(\infty) +{\cal O}(1) \sim \tau$ for $\tau\gg 1$.
Consequently, the purity factor $\chi$ asymptotically goes to zero as
$\tau^{-1/2}$.

\section{Photon distribution in the Gaussian state}\label{ap-3}

In the most compact form the information on the photon distribution
$f(n)$ in some mode (we suppress here the mode index) is contained in
the {\it generating function\/}
\[
G(z)=\sum_{n=0}^{\infty}\,f(n)z^n
\]
For the most general Gaussian state it was given in \cite{1mod,DM94}
(for a single mode):
\be
G(z)=[{\cal G}(z)]^{-1/2}\exp\left(\frac1{D}\left[
\frac{zg_1 -z^2 g_2}{{\cal G}(z)} - g_0\right] \right)
\label{GF}
\ee
where
\be
{\cal G}(z)=\frac14\left[(1+z)^2 +4\left(UV-Y^2\right) (1-z)^2 +
2(U+V)\left(1-z^2\right)\right]
\label{calG}
\ee
\[
D=1+2(U+V) +4 \left(UV-Y^2\right) = 4{\cal G}(0)
\]
\[
g_0= \langle \hat{p}\rangle^2(2U+1) +\langle \hat{q}\rangle^2(2V+1)
-4\langle \hat{p}\rangle\langle \hat{q}\rangle Y
\]
\begin{eqnarray*}
g_1&=& 2\langle \hat{p}\rangle^2\left(U^2 +Y^2 +U +\frac14\right)
+2\langle \hat{q}\rangle^2\left(V^2 +Y^2 +V +\frac14\right)\\
&&-4\langle \hat{p}\rangle\langle \hat{q}\rangle Y(U+V+1)
\end{eqnarray*}
\[
g_2= 2\langle \hat{p}\rangle^2\left(U^2 +Y^2 -\frac14\right)
+2\langle \hat{q}\rangle^2\left(V^2 +Y^2 -\frac14\right)
-4\langle \hat{p}\rangle\langle \hat{q}\rangle Y(U+V)
\]
If $\langle \hat{p}\rangle=\langle \hat{q}\rangle=0$,
then the probability $f(n)$ is
expressed in terms of the Legendre polynomials: see equation (\ref{fotdis}).
In the generic case $f(n)$ is related to the two-dimensional `diagonal'
Hermite polynomials \cite{1mod}:
\be
f(n)=\frac{{\cal F}_0}{n!}H_{nn}^{\{{\cal R}\}}\left( x, x^*\right)
\label{f-Hnn}
\ee
where
\[
{\cal F}_0=f(0)= 2D^{-1/2}\exp\left(-g_0/D\right)
\]
\[
x=\frac{\sqrt{2}\left\{(2V-1)\langle \hat{q}\rangle -2Y\langle \hat{p}\rangle
+i\left[(1-2U)\langle \hat{p}\rangle +2Y\langle \hat{q}\rangle\right]\right\}}
{2(U+V) -4\left(UV-Y^2\right) -1}
\]
and $2\times2$ symmetric matrix ${\cal R}$ has the elements
\[
{\cal R}_{11}={\cal R}_{22}^* = \frac2{D}(V-U-2iY), \quad
{\cal R}_{12}={\cal R}_{21} =\frac1{D}\left[1-4\left(UV-Y^2\right)\right]
\]
The two-dimensional Hermite polynomials are defined via  the
expansion \cite{B}
\be
\exp\left(-\frac12 {\bf a}{\cal R}{\bf a} +{\bf a}{\cal R}{\bf x}\right)
=\sum_{m,n=0}^{\infty} \frac{a_1^m a_2^n}{m!n!}
H_{mn}^{\{{\cal R}\}}\left( x_1, x_2\right)
\label{def-herm}
\ee
where $ {\bf x}=\left( x_1, x_2\right)$, ${\bf a}=\left( a_1, a_2\right)$.
The properties of these polynomials were studied recently in
\cite{DM94,D94}. In particular, they can be expressed as finite sums of the
products of the usual (one-dimensional) Hermite polynomials. The
corresponding formula for the probabilities reads \cite{1mod}
\be
f(n)={\cal F}_0\left(\frac{\Delta}{D}\right)^n
\sum_{k=0}^n \left(\frac{S}{\Delta}\right)^k \frac{n!}{[(n-k)!]^2 k!}
\left|H_{n-k}(\xi)\right|^2
\label{f-Herm}
\ee
where
\[
\Delta=\sqrt{(U-V)^2 +4Y^2}, \quad S=4\left(UV-Y^2\right)-1
\]
\[
\xi=\frac{(2V+1)\langle \hat{q}\rangle -2Y\langle \hat{p}\rangle
+i\left[(1+2U)\langle \hat{p}\rangle -2Y\langle \hat{q}\rangle\right]}
{\left[2D(V-U-2iY)\right]^{1/2}}
\]

The photon distribution function (\ref{fotdis})
 can be simplified
in the long-time limit $\tau\gg 1$,
when the average number of created photons
${\cal N}\equiv \bar{n}\approx (V+U)/2$ exceeds $1$.
Then the mean-square fluctuation of the photon number 
has the same
order of magnitude as the mean photon number itself, 
$\sqrt{\sigma_n}\approx \sqrt2 {\cal N}$, and
 the most significant part
of the spectrum corresponds to the values $n\gg 1$.
Using the Laplace--Heine asymptotical formula for the
Legendre polynomial \cite{Szego}
\[
P_n(z)\approx \frac{\left(z+\sqrt{z^2-1}\right)^{n+1/2}}
{\sqrt{2\pi n}\left(z^2-1\right)^{1/4}}\,, \quad n\gg 1
\]
one can simplify (\ref{fotdis}) for the fixed values of the invariant
variances $u$ and $v$ as
\begin{equation}
f(n) \approx \frac1{\sqrt{\pi n(v-u)}}
\left(\frac{2v-1}{2v+1}\right)^{n+1/2}
\label{asf}
\end{equation}
provided the positive difference $v-u$ is not too small.
Another approximate formula can be used if $v\gg 1$ but
$u\sim 1$:
\be
f(n)\approx \frac{\sqrt2(2u-1)^{n/2}}
{\sqrt{v}(2u+1)^{(n+1)/2}} e^{-n/(2v)}
P_n\left(\frac {2u}{\sqrt{4u^2-1}}\right),
\quad n\ll 8v^2.
\label{aprdis}
\end{equation}

\Figures

\Figure{
The minimal variance $u_1$, the maximal variance $v_1$,
the purity factor $\chi$,
the mean photon number ${\cal N}_1$  and  Mandel's parameter $Q$
of the fundamental mode $\mu=1$ versus the universal parameter $\kappa$
(\ref{defpar}).
\label{fig-1}     }

\Figure{
The photon distribution function in the fundamental field mode
$\mu=1$ for $\tau=5$ and $\gamma=0$, $p=2$
(points connected with solid lines).
Points connected with dashed lines correspond to the Planck distribution
with the same mean photon number.
\label{fig-comp}     }

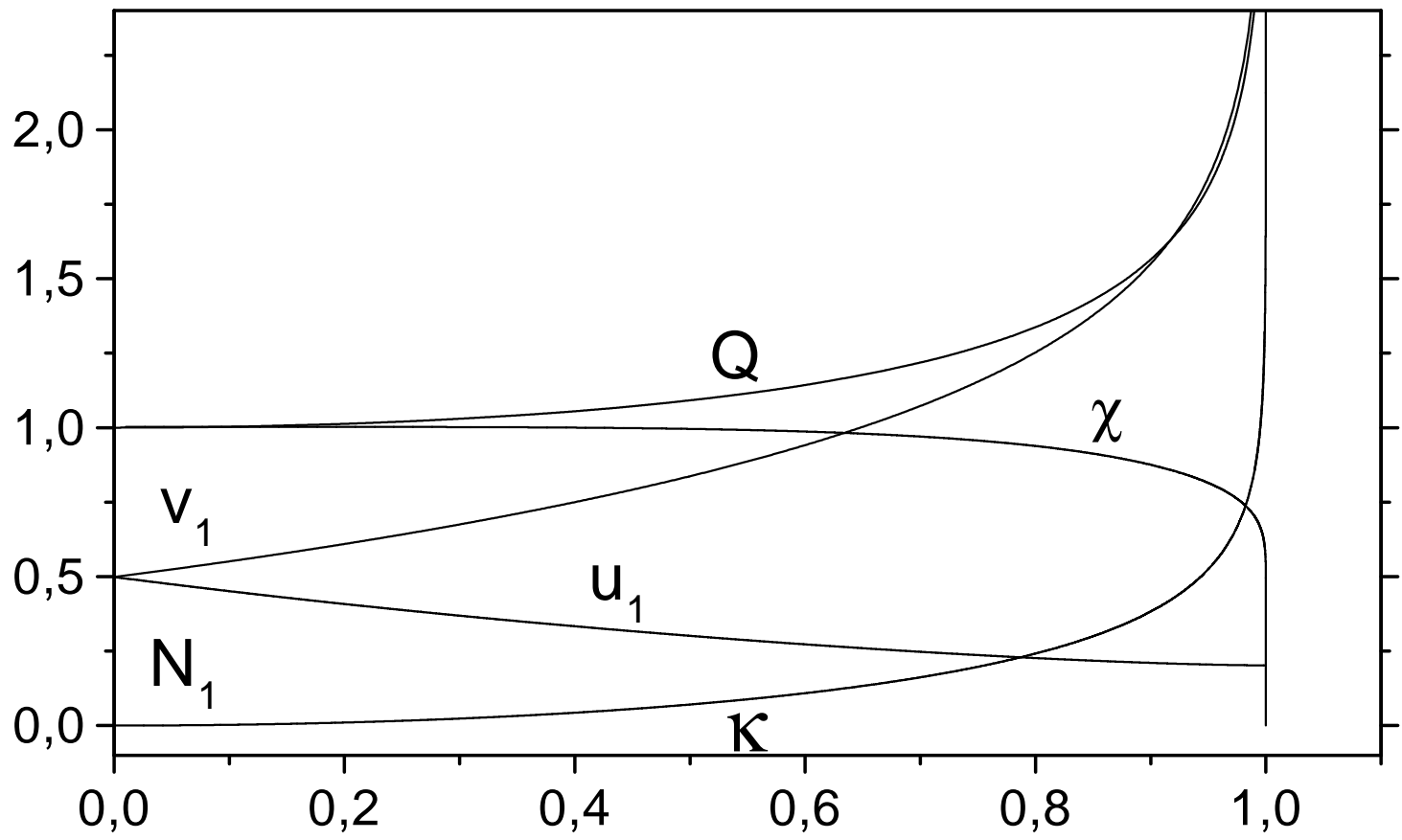
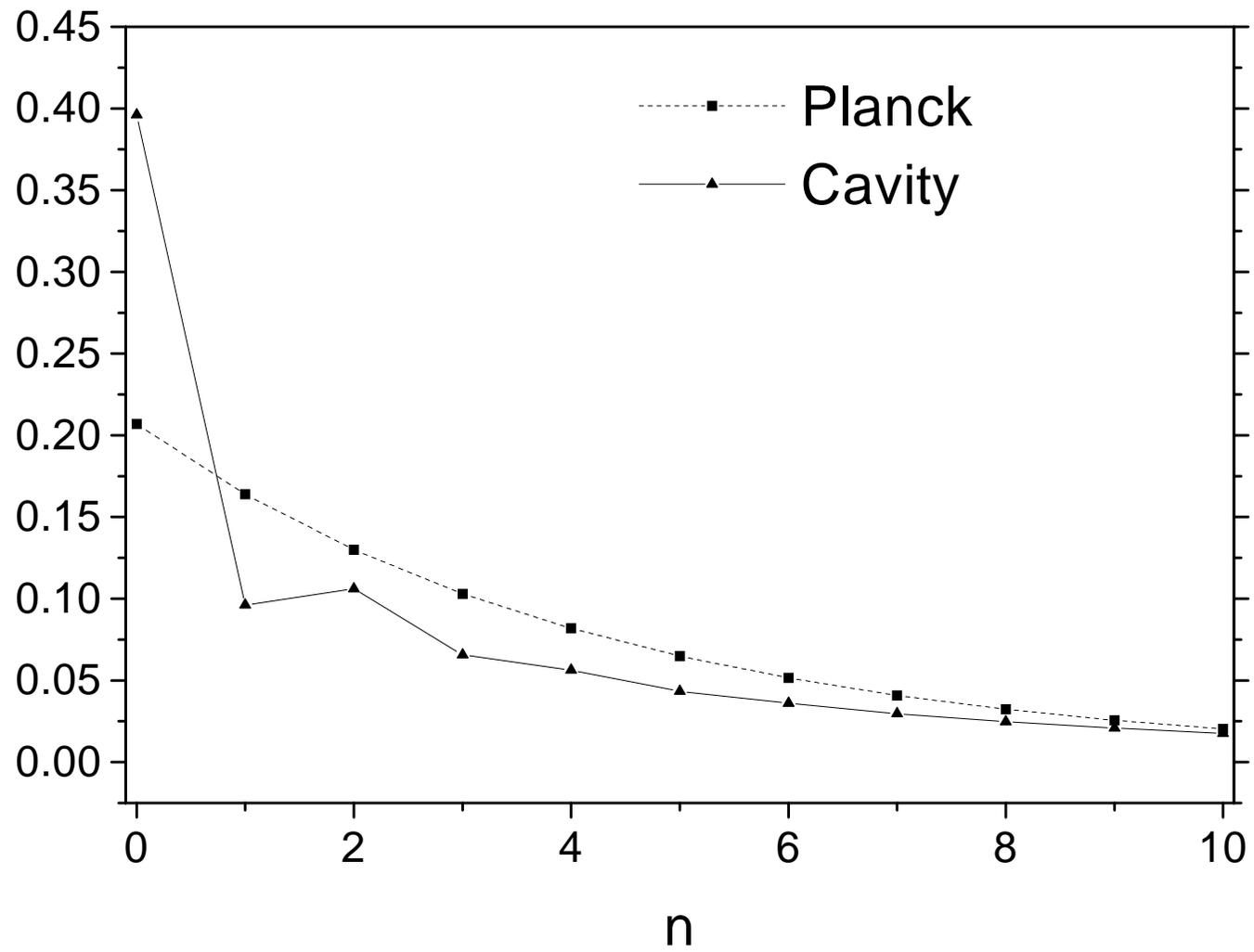
\end{document}